Hetero-epitaxial EuO Interfaces Studied by Analytic Electron Microscopy

Julia A. Mundy,[1,a] Daniel Hodash,[2,a] Alexander Melville,[2] Rainer Held,[2] Thomas Mairoser,[3] David A. Muller,[1,4] Lena F. Kourkoutis,[1,4] Andreas Schmehl,[3] and Darrell G. Schlom[2,4]

[1]*School of Applied and Engineering Physics, Cornell University, Ithaca, New York 14853, USA*
[2]*Department of Materials Science and Engineering, Cornell University, Ithaca, New York 14853, USA*
[3]*Zentrum für Elektronische Korrelationen und Magnetismus, Universität Augsburg, Universitätsstraße 1, D-86159 Augsburg, Germany*
[4]*Kavli Institute at Cornell for Nanoscale Science, Ithaca, New York 14853, USA*

With nearly complete spin polarization, the ferromagnetic semiconductor europium monoxide could enable next-generation spintronic devices by providing efficient ohmic spin injection into silicon. Spin injection is greatly affected by the quality of the interface between the injector and silicon. Here, we use atomic-resolution scanning transmission electron microscopy in conjunction with electron energy loss spectroscopy to directly image and chemically characterize a series of EuO|Si and EuO|YAlO$_3$ interfaces fabricated using different growth conditions. We identify the presence of europium silicides and regions of disorder at the EuO|Si interfaces, imperfections that could significantly reduce spin injection efficiencies via spin-flip scattering.

---

[a] These authors contributed equally to this work.

The ferromagnetic semiconductor EuO is a promising candidate material for next-generation spintronics. Not only does the half-metal EuO exhibit a spin polarization of 96%,[1–3] but it can also be conductance-matched to desired substrates by doping with trivalent rare-earths such as lanthanum.[4] Importantly, doped EuO retains the high spin-polarization of the undoped material,[4] possibly providing a solution to the conductivity mismatch problem of metallic spin injector materials.[2,4,5] Furthermore, EuO is the only magnetic binary oxide that might be stable in direct contact with the pervasive long spin-lifetime semiconductor, silicon.[6] EuO has a large magnetic moment of 7 $\mu_B$ per europium atom,[7] exceptional magneto-optical properties, with a demonstrated Faraday rotation of $8.5 \times 10^5$ deg/cm in a magnetic field of 2 T,[8] and a metal-insulator transition with a resistance change that can exceed 13 orders of magnitude.[9,10] While undoped EuO has a low Curie temperature of 69 K, chemical doping can increase the Curie temperature ($T_c$) significantly[1,11–13] and biaxial strain can further manipulate $T_c$.[14,15]

Multiple methods are proposed for injecting polarized spin currents into a channel material including Schottky barrier injection,[16] tunneling,[17,18] ohmic injection,[5] and hot electron injection.[16] In principle, ohmic injection should be best suited for technological applications as it allows both large current and highly efficient spin injection. EuO could be an efficient ohmic injector material for silicon as it can preserve a high spin-polarization at a contact as demonstrated with Andreev reflection[1,4] and can be conductance-matched[5] to silicon.[4] The EuO|Si interface, however, must be optimized to prevent spin-flip scattering from impurity phases. Even if the Eu|Si interface itself is thermodynamically stable, $SiO_2$ may form during growth if the environment is too oxidizing. Likewise, $EuSi_2$ may form if the growth environment is too reducing.[19]

Achieving conditions within such a narrow growth window is an immense challenge common to the epitaxial integration of all oxides with silicon.[20,21] Moreover, the +5.6% lattice mismatch[22] between EuO and Si could lead to misfit dislocations or other disorder. While scanning transmission electron microscopy (STEM) is known as an effective characterization tool for site-specific imaging of defects and secondary phases, the extreme water and oxygen sensitivity of EuO renders the sample preparation difficult. Here, we use water-free sample preparation to achieve atomic-resolution images and spectroscopic analysis of EuO|Si interfaces fabricated under a variety of growth conditions. We quantify the presence of an interfacial Eu-Si phase and investigate the spatial extent of interfacial disorder.

The EuO|Si interface can be analyzed with a variety of signals accessible on the electron microscope. Atomic-resolution high-angle annular dark field (HAADF) STEM images provide a qualitative atomic number ($Z$) contrast and as such can be used to distinguish heavy europium ($Z = 63$) and europium-containing phases such as europium silicides from silicon ($Z = 14$) columns. Chemical analysis from electron energy loss spectroscopy (EELS) can further confirm the elemental compositions and bonding information. For the core-loss transitions, the energy-loss near-edge fine-structure (ELNES) correlates with the local density of states above the Fermi level, specific to the site, element, and angular momentum.[23] Analysis of the EELS line shape can thus produce atomic-resolution information about the oxidation state and bonding information of the constituent species present.[24–27] While other techniques including X-ray absorption spectroscopy[4,12,28] and hard X-ray photoemission spectroscopy[29] have detected the presence of multiple europium valence states in $EuO_{1\pm\delta}$ films, electron microscopy is

uniquely poised to investigate variations in the interface structure at the atomic scale, which are not discernable by bulk characterization. Transmission electron microscopy on EuO is challenging due to its reactivity with air and we are aware of only one report of its use,[4] though no images were shown in that report. Here we show images. We also note that the *in situ* scanning tunneling microscopy of EuO films was recently achieved.[30]

Thin films of La-doped EuO were grown on (001) Si and on (110) YAlO$_3$ by molecular-beam epitaxy (MBE) in Veeco 930 and GEN 10 MBE systems. The fluxes of evaporated europium and lanthanum were calibrated using a quartz crystal microbalance. For low-temperature flux-matched conditions, the oxygen flux was meticulously calibrated by analyzing reflection high-energy electron diffraction (RHEED) patterns often in combination with a mass spectrometer. For growth in the adsorption-controlled regime (temperatures above ~400°C), where excess atoms of the volatile europium metal evaporate, a sufficiently low oxygen flux was employed as the limiting reagent to achieve stoichiometric EuO.[31] The grown films were capped *in situ* with either aluminum or amorphous silicon (a-Si) before removing them from the MBE system to protect the EuO from further oxidation in air. The structural quality was probed by X-ray diffraction (XRD). Sample growth details and XRD patterns are given in Supplemental Discussion 1 and Supplemental Figure S1 respectively.[32] Cross-sectional TEM specimens were prepared using the FEI Strata 400 focused ion beam with final ion milling performed at 2 keV to minimize surface damage as well as mechanical polishing with water-free solvents followed by a low-angle, low-energy ion milling cleaning step. HAADF-STEM imaging and EELS line scans were performed on a 200 keV FEI Technai F-20, with a 1.6 Å STEM probe size and an EELS energy resolution of 0.7 eV. A 100 keV 5$^{th}$-order aberration-corrected Nion UltraSTEM was also used to acquire HAADF-STEM images and EELS spectroscopic images with a probe size of 1 Å, an energy resolution of 0.6 eV, and nearly 200 pA of useable beam current. Simultaneous line profiles of Eu-$N_{4,5}$ and Si-$L_{2,3}$ edges across the interface were acquired with a 0.2 eV/channel dispersion.

The relatively low lattice mismatch and low reactivity[4] between EuO and the YAlO$_3$ substrate provides us with an experimental basis for HAADF-STEM images of high-quality epitaxial EuO films. With a lattice parameter of 5.1426 Å,[22] (001) EuO aligns with the (110) YAlO$_3$ surface with [110] EuO || [001] YAlO$_3$, resulting in an average mismatch strain of 1.8%.[4] A high-resolution HAADF-STEM image of an undoped EuO film grown on YAlO$_3$ in the adsorption-controlled regime[31] is shown in Fig. 1(a). High- and low-magnification images of a 5% La-doped EuO film exhibiting a $T_C$ of 109 K grown by flux-matching are shown in Figs. 1(b) and 1(c), respectively. In both growth regimes, the HAADF-STEM images indicate that the films are clearly epitaxial. A two-dimensional EELS spectroscopic map of the interface of the latter film, Supplemental Figure S2,[30] shows the presence of a chemically abrupt interface without interfacial europium valence changes. This suggests that the approximately two-monolayer dark region observed in Fig. 1(b) could be attributed to a dechanneling effect.[33] Both growth techniques produced chemically abrupt, coherent EuO|YAlO$_3$ interfaces.

In contrast to growth on YAlO$_3$, epitaxy of EuO on the technologically relevant substrate silicon is challenging by the propensity of silicon interfaces to form SiO$_2$ and silicides,[6,20,34,35] and by the much larger lattice mismatch of +5.6 % between (001) EuO and (001) Si. Figure 2 shows HAADF-STEM images of an undoped EuO film grown on a bare (001) Si substrate. In contrast to the chemically abrupt interfaces between EuO and YAlO$_3$, a 2 nm thick disordered layer can be clearly detected between EuO and silicon. This disordered layer is less crystalline than the substrate and the remainder of the EuO film. The XRD pattern, shown in Supplemental Figure S1, shows the presence of a secondary phase that can be assigned to EuSi$_2$.[32] As shown in Fig. 2(c), a distinct additional crystalline phase is present in some regions of interface. Single crystal diffraction simulations oriented along the [010] zone axis of EuSi$_2$ match the diffractogram (inset to Fig. 2(c)) of the corresponding region of the STEM image.

EELS was used to reveal the chemistry of the interface, including the impurity phases, as shown in Fig. 3. Figure 3(a) shows that two distinct EELS fingerprints can be extracted from the Eu-$N_{4,5}$ edge, one from the secondary phase observed at the interface, matching the bulk film, and the other from the disordered interface region. The corresponding fingerprints from the Eu-$M_{4,5}$ edge are shown in Fig. 3(b). The spectroscopic signal present in both the precipitates at the interface and the bulk film can be assumed to be Eu$^{2+}$ since magnetic measurements of all films considered here can only match the highly magnetic EuO. This is also the expected valence state for EuSi$_2$. which corroborates the structural assignment of the interfacial phase to EuSi$_2$ from XRD and the diffractogram from the STEM image. Comparison with Eu-$M_{4,5}$ reference spectra[36] confirms the Eu$^{2+}$ valence state. The signal from the disordered interfacial region above the EuSi$_2$ shows an increase in the onset energy, consistent with an Eu$^{2+}$ to Eu$^{3+}$ bonding state transition.[36] EELS line spectra from the region shown in Figs. 3(c) and 2(c) simultaneously capture the Si-$L_{2,3}$ and Eu-$N_{4,5}$ edges. A multivariate curve resolution analysis of the Eu-$N_{4,5}$ fine structure separated the lower valence Eu$^{2+}$ signal from the Eu-

$^{3+}$ signal as shown in Fig. 3(d). The normalized concentration of silicon is calculated from the Si-$L_{2,3}$ edge and overlaid on the two distinct signals from the europium edge in Fig. 3(d). The EELS concentrations extracted track the intensities of the HAADF image, with silicon prevalent in the substrate and through the EuSi$_2$, Eu$^{2+}$ in EuSi$_2$ impurities and the film, and the Eu$^{3+}$ in the disordered region at the interface.

The observation of a europium silicide impurity phase as well as the Eu$^{3+}$ signal at the EuO|Si interface highlights the difficulty in abruptly transitioning from silicon to EuO during growth. If the initial growth conditions are too oxidizing, the silicon substrate will be oxidized to amorphous SiO$_2$ and the crystalline template will be lost.[37] On the other hand, if the initial growth conditions are insufficiently oxidizing, europium metal will come into contact with the silicon substrate and europium silicide will form, as was the case for the film analyzed in Fig. 2. It is natural to err on the oxygen deficient side as an epitaxial EuO film can be achieved despite the presence of the unwanted europium silicide reaction layer. Examples abound in the literature of such insufficiently oxidizing conditions being used in the epitaxial growth of EuO and other oxides on silicon and the silicide reactions that result.[20,29,34,38,39]

In an effort to prevent the formation of EuSi$_2$ and the disordered Eu$^{3+}$ layer observed in Fig. 3, SrO was deposited on the silicon substrate before growth of additional EuO films. SrO is known to be able to provide a template on the silicon surface suitable for the growth of EuO[4,37] and might be stable in direct contact with silicon.[6] A sample with two monolayers of SrO and a sample with five monolayers of SrO were grown to systematically study the evolution of the interfacial structure. Contrary to what we observed for the EuO film grown directly on Si, there is no evidence of EuSi$_2$ in XRD patterns recorded on either sample (Supplemental Figure S1)[30]. Figure 4 shows HAADF-STEM images and chemical analysis of the interfaces. The film with two monolayers of SrO is shown in Fig. 4(a). A high-magnification image in Fig. 4(b) shows that a ~2 nm thick disordered region is still present above the SrO layer. In the region shown in Fig. 4(b) a phase that appears to be similar to the $EuSi_2$ phase from the previous sample is observed. EELS analysis confirms that europium is present in this layer with a $Eu^{2+}$ oxidation state, suggesting that the interfacial regions with large bright spots contain europium silicides, below the detection limit of XRD. Possible reasons for the formation of such a reaction phase include islanding of the SrO buffer layer, (which is two monolayers thick on average) leading to incomplete coverage of the silicon substrate or the diffusion of europium through the SrO layer. An additional film with five-monolayers of SrO inserted at the interface is shown in Fig. 4(c) and Fig. 4(d). While there is less disorder observed at the interface, the EELS analysis (Fig. 4(e)) demonstrates that there is still $Eu^{3+}$ present for approximately 5 nm above the interface. Moreover, there is an about 1-2 nm thick region for which $Eu^{2+}$ and silicon are present, evidence of a small amount of europium silicide.

In summary, we have grown epitaxial EuO films on silicon and $YAlO_3$ substrates and imaged their microstructure. Although standard XRD techniques cannot always detect the presence of additional phases, HAADF-STEM images clearly show the existence of phases not visually matching either EuO or silicon in the vicinity of the interface. Combined with analysis of EELS peak shifts, the presence of a europium silicide phase identified to be $EuSi_2$ was detected as well as an interfacial increase in the europium valence. Although the deposition of a five-monolayer-thick SrO buffer layer did not fully prevent the formation of unwanted interfacial valence changes to the europium, an optimized SrO layer and/or a further increase in the SrO layer thickness may create a more homogeneous interface without holes that is suitable for spin injection by tunneling. For efficient ohmic spin injection, the EuO|Si interface needs further optimization because the formation of unwanted phases can interfere with precise conductance-matching and impurities may lead to spin-flip scattering. The ability of combined HAADF-STEM and EELS to provide atomic resolution information about the structure of EuO|Si interfaces and identify subtle valence changes near them below the detection limit of X-ray analysis makes these techniques ideal for evaluation and guiding improvements of such interfaces.

The authors acknowledge useful conversations with Daniel E. Shai and Stephen Dacek. The work at Cornell was supported by the AFOSR (Grant No. FA9550-10-1-0123). This work made use of the electron microscopy facility of the Cornell Center for Materials Research (CCMR) with support from the National Science Foundation Materials Research Science and Engineering Centers (MRSEC) program (DMR 1120296) and NSF IMR-0417392. The work in Augsburg was supported by the DFG (Grant No. TRR 80) and the EC (oxIDes). J.A.M. acknowledges financial support from the Army Research Office in the form of a National Defense Science & Engineering Graduate Fellowship and from the National Science Foundation in the form of a graduate research fellowship. A.J.M. acknowledges support from the NSF IGERT program (NSF Award DGE-

**Figures**

**Figure 1** – Cross-sectional HAADF-STEM images of epitaxial EuO films grown on YAlO$_3$, viewed along the [110] zone axis of the EuO film and the [1$\bar{1}$0] zone axis of the YAlO$_3$ substrate. (a) Undoped EuO grown by adsorption control at 400°C and (b) ~5% La-doped EuO grown under flux-matched conditions at 250°C. (c) Low-magnification view of (b) depicting the entire film. The interfaces are coherent and chemically abrupt.

**Figure 2** – Cross-sectional HAADF-STEM images of epitaxial EuO films grown on silicon at 350°C, viewed along the [110] zone axis of both the EuO film and silicon substrate. (a) Low-magnification view of the whole film. (b) and (c) present close-ups of different regions of the silicon-EuO interface. (b) Shows a region of the sample exhibiting a comparatively uniform interface, yet there is an approximately 2 nm thick disordered region between the EuO and silicon substrate. (c) Shows a region containing a ~5 nm thick crystalline europium silicide precipitate that was further analyzed with EELS (Fig. 3). A diffractogram of the impurity phase is shown as an inset.

**Figure 3** – EELS fine structure analysis of the EuO|Si interface. Distinct EELS fingerprints corresponding to Eu$^{2+}$ and Eu$^{3+}$ shown for the Eu $N_{4,5}$-edge and Eu $M_{4,5}$-edge are shown in (a) and (b), respectively. The concentrations of Eu$^{2+}$, Eu$^{3+}$, and Si through the region in the HAADF-STEM image shown in (c) are plotted in (d). The precipitate identified in Fig. 2 and shown in (c) is identified as having Eu$^{2+}$ and silicon present, consistent with europium silicide. Above the europium silicide, a distinct Eu$^{3+}$ is identified before the onset of the EuO.

**Figure 4** – Cross-sectional HAADF-STEM images of epitaxial EuO films grown on silicon with a SrO buffer layer in between, viewed along the [110] zone axis of both the EuO film and silicon substrate. Low-magnification and high-magnification images of a film with a 2 ML SrO buffer layer are shown in (a) and (b), respectively. Low-magnification and high-magnification images of a film with a 5 ML SrO buffer layer are shown in (c) and (d), respectively. EELS analysis of the interface shown in (d) is presented in (e), demonstrating the persistence of both Eu$^{3+}$ and evidence for europium silicide at the interface despite the thick SrO buffer layer.

**Figure 1**

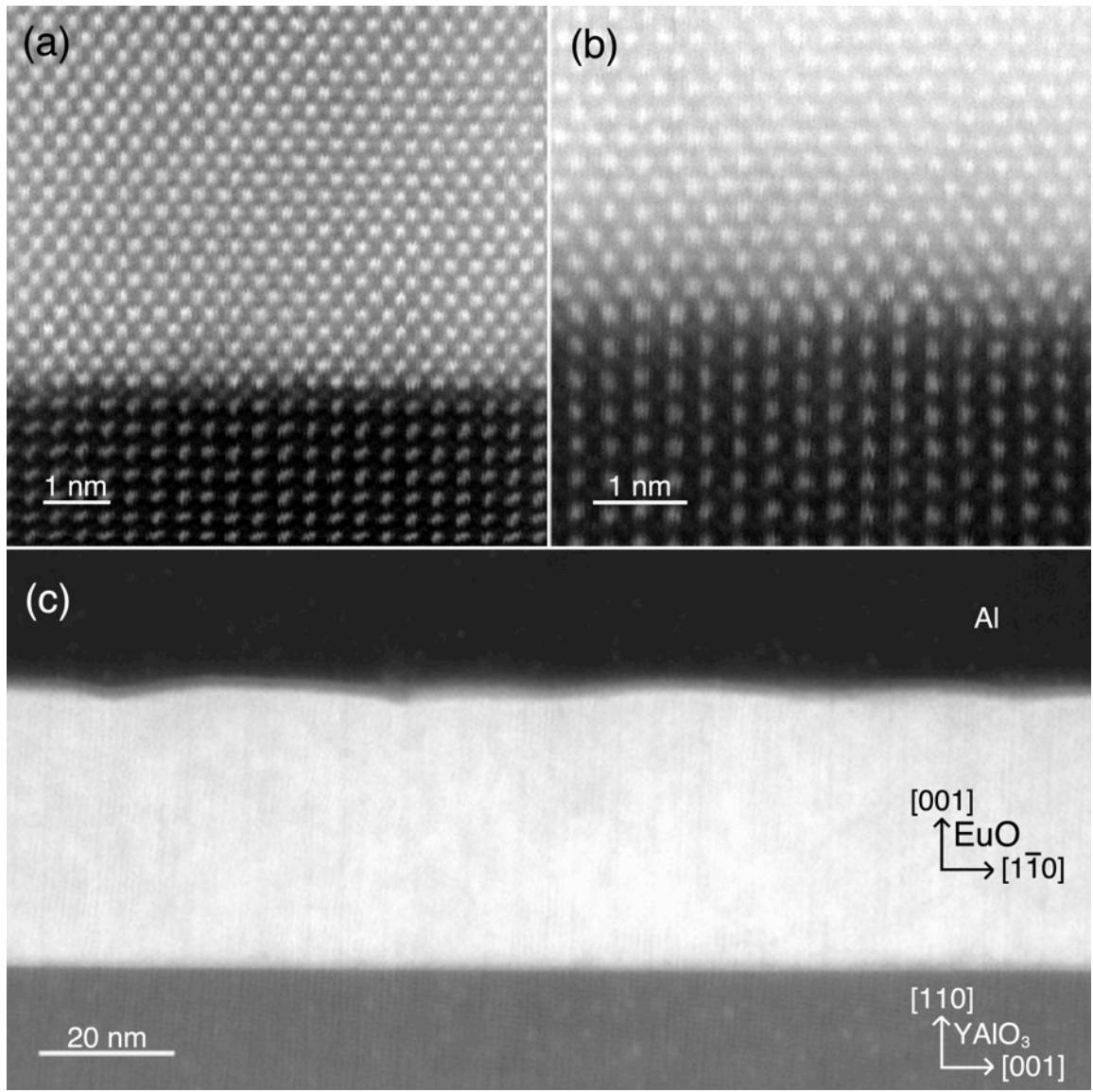

**Figure 2**

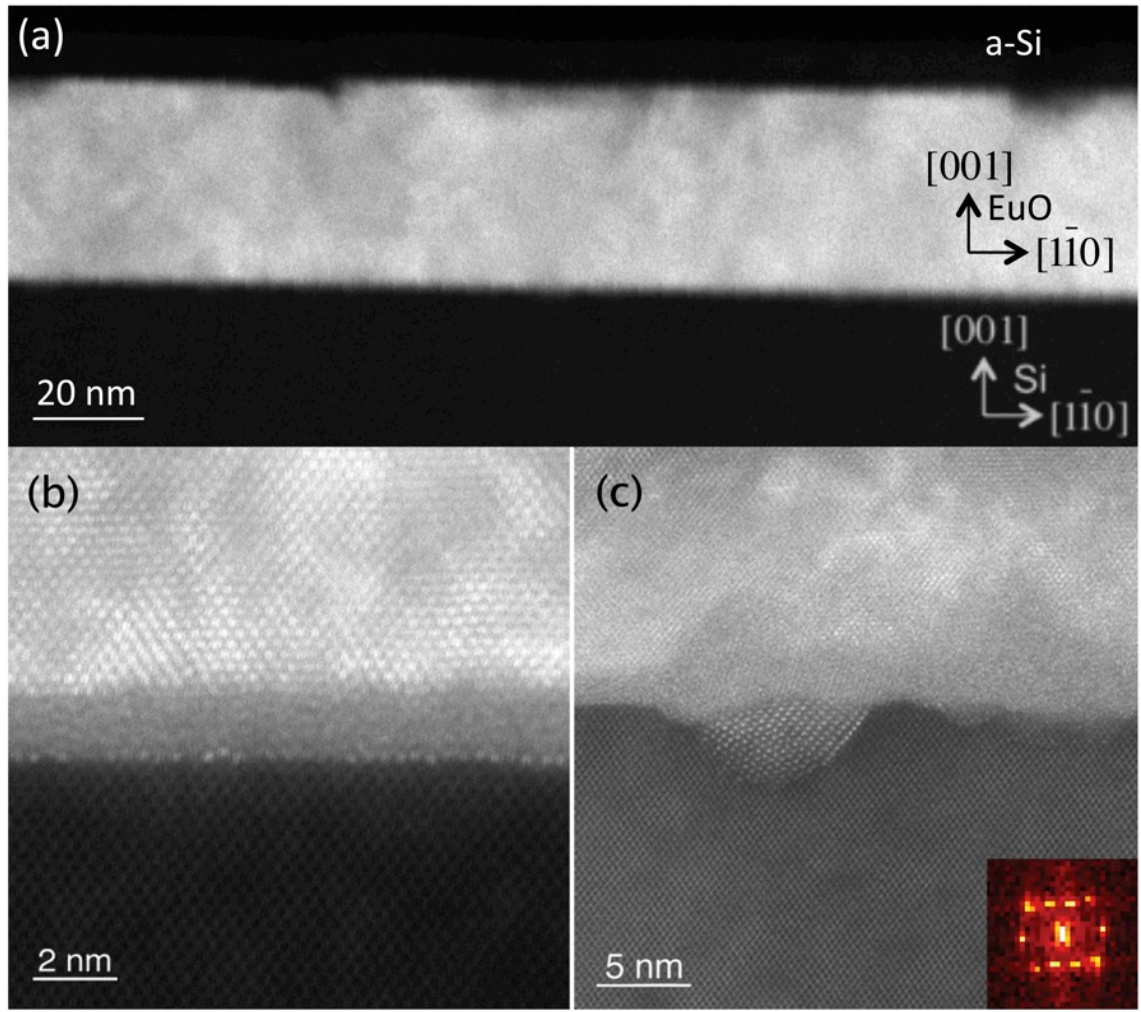

**Figure 3**

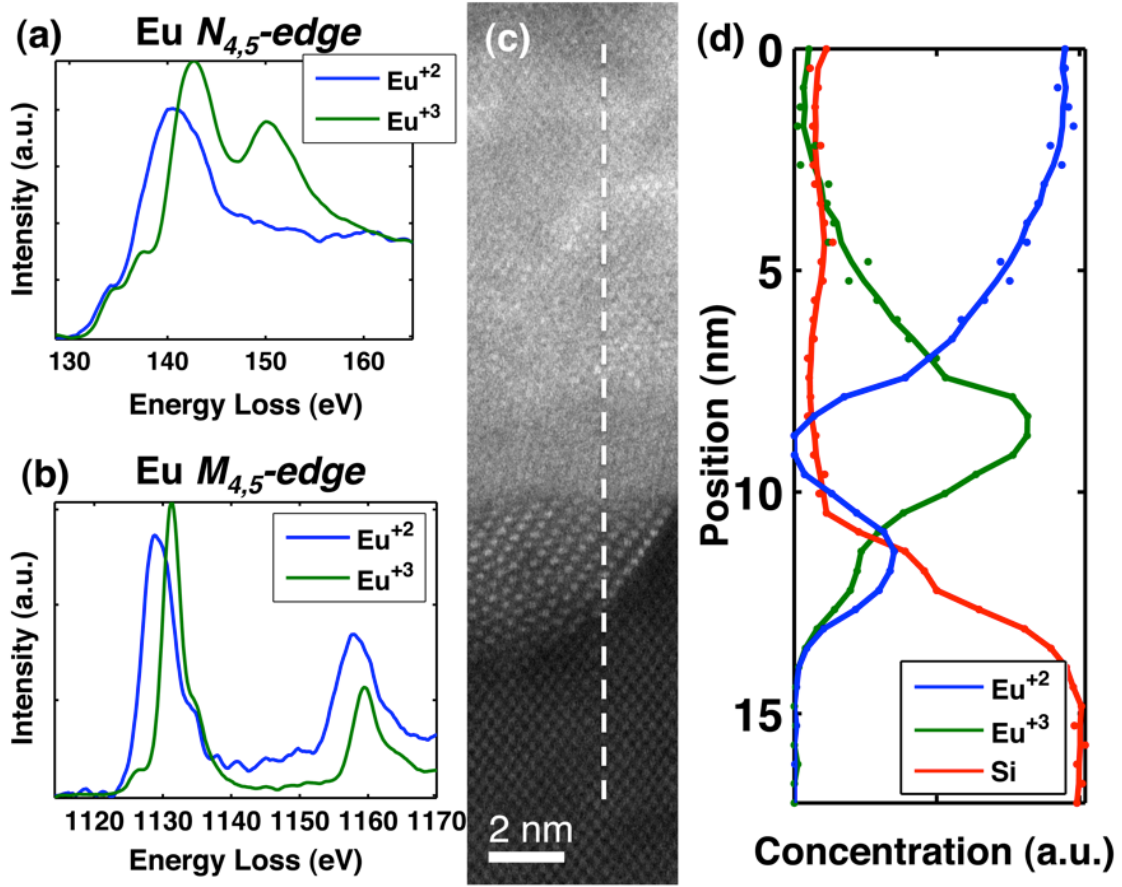

**Figure 4**

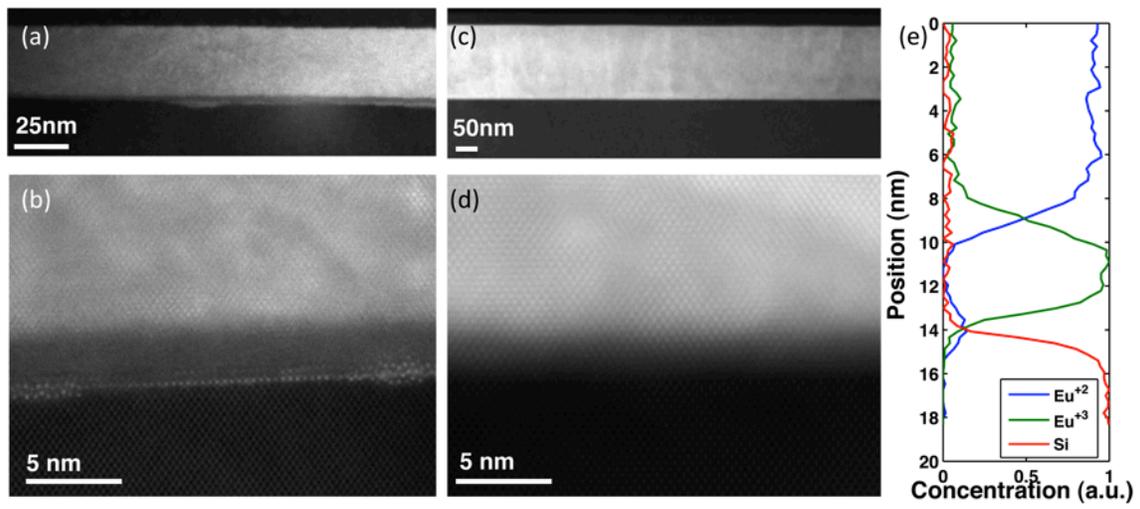